\documentclass[final,5p,times,twocolumn]{elsarticle}

\journal{Physics Letters B}

\usepackage{color}
\begin{document}

\begin{frontmatter}

\title{Measurement of polarisation observables in $K^0_s\Sigma^+$
       photoproduction off the proton}




{\author[Bonn]{R.~Ewald\fnref{label1}}
\author[HISKP,PNPI]{A.V.~Anisovich}
\author[Bonn]{B.~Bantes}
\author[HISKP]{O.~Bartholomy}
\author[HISKP,PNPI]{D.~Bayadilov}
\author[HISKP]{R.~Beck}
\author[PNPI]{Y.A.~Beloglazov}
\author[Giessen]{K.-T.~Brinkmann}
\author[Florida]{V.~Crede} 
\author[Bonn]{H.~Dutz}
\author[Bonn]{D. Elsner}
\author[Bonn]{K.~Fornet-Ponse}
\author[Bonn]{F.~Frommberger}
\author[HISKP]{Ch.~Funke}
\author[PNPI]{A.~B.~Gridnev}
\author[Giessen]{E.~Gutz}
\author[Bonn]{J.~Hannappel}
\author[Bonn]{W.~Hillert}
\author[HISKP]{P.~Hoffmeister}
\author[Basel]{I.~Jaegle}
\author[Bonn]{O.~Jahn}
\author[Bonn]{T.C~Jude\corref{cor1}}
\author[HISKP]{J.~Junkersfeld}
\author[HISKP]{H.~Kalinowsky}
\author[Bonn]{S.~Kammer}
\author[Bonn]{V.~Kleber\fnref{label3}}
\author[Bonn]{Frank Klein}
\author[Bonn]{Friedrich Klein}
\author[HISKP]{E.~Klempt}
\author[Basel]{B.~Krusche}
\author[HISKP]{M.~Lang}
\author[Groningen]{H.~L\"ohner}
\author[PNPI]{I.V.~Lopatin}
\author[Bonn]{D.~Menze}
\author[Giessen]{T.~Mertens}
\author[Groningen]{J.G.~Messchendorp}
\author[Giessen]{V.~Metag}
\author[Giessen]{M.~Nanova}
\author[HISKP,PNPI]{V.A.~Nikonov}
\author[HISKP,PNPI]{D.~Novinski}
\author[Giessen]{R.~Novotny}
\author[Bonn]{M.~Ostrick\fnref{label4}}
\author[Groningen]{L.~Pant\fnref{label5}}
\author[HISKP]{H.~van Pee}
\author[Groningen]{A.~Roy\fnref{label6}}
\author[HISKP,PNPI]{A.V.~Sarantsev}
\author[Groningen]{S.~Schadmand\fnref{label7}}
\author[HISKP]{C.~Schmidt}
\author[Bonn]{H.~Schmieden}
\author[Bonn]{B.~Schoch}
\author[Groningen]{S.~Shende}
\author[HISKP]{V.~Sokhoyan}
\author[Bonn]{A.~S{\"u}le}
\author[PNPI]{V.V.~Sumachev}
\author[HISKP]{T.~Szczepanek}
\author[HISKP]{U.~Thoma}
\author[Giessen]{D.~Trnka}
\author[Groningen]{R.~Varma}
\author[HISKP]{D.~Walther}
\author[HISKP]{Ch.~Wendel}                    
\address[Bonn]{Physikalisches Institut, Universit{\"a}t Bonn, Germany}
\address[HISKP]{Helmholtz Institut f\"ur Strahlen- und Kernphysik, Universit{\"a}t Bonn, Germany}
\address[PNPI]{Petersburg Nuclear Physics Institute, Gatchina, Russia}
\address[Basel]{Departement Physik, Universit\"at Basel, Switzerland}
\address[Florida]{Department of Physics, Florida State University, Tallahassee, USA}
\address[Giessen]{II. Physikalisches Institut, Universit\"at Gie{\ss}en, Germany}
\address[Groningen]{Kernfysisch Versneller Instituut, Groningen, The Netherlands}
\fntext[label1]{Now at DLR, Cologne, Germany}
\cortext[cor1]{Corresponding author, jude@physik.uni-bonn.de}
\fntext[label3]{Now at German Research School for Simulation Sciences, J\"ulich, Germany}
\fntext[label4]{Now at Institut f\"ur Kernphysik, Universit\"at Mainz, Germany}
\fntext[label5]{On leave from Nucl. Phys. Division, BARC, Mumbai, India}
\fntext[label6]{On leave from Department of Physics, IIT, Mumbai, India}
\fntext[label7]{Present address: Institut f\"ur Kernphysik and J\"ulich Center for Hadron Physics,
Forschungszentrum J\"ulich, Germany}
}
%
%
%

\begin{abstract}
The reaction $\gamma \, p \rightarrow K^0_S\,\Sigma^+$
is studied in the photon energy range from threshold.
  Linearly polarised photon beams from coherent bremsstrahlung 
enabled the first measurement of photon beam asymmetries in this reaction up to $E_\gamma = 2250$\,MeV.
In addition, the recoil hyperon polarisation was determined through the 
asymmetry in the weak decay $\Sigma^+ \rightarrow p \pi^0$ up to $E_\gamma = 1650$\,MeV.
The data are compared to partial wave analyses, and the possible
impact on the interpretation of a recently observed cusp-like structure
near the $K^*$ thresholds is discussed.
\end{abstract}


\end{frontmatter}

\section{Introduction}
\label{intro}
The CBELSA/TAPS experiment at the Electron Stretcher Accelerator
ELSA is devoted to the investigation of the
structure of the nucleon at low energies.
While the high energy and associated short distance dynamics are 
well understood 
and put into the commonly accepted frame of quantum chromodynamics (QCD), 
our knowledge is still rather limited at the size/mass scale of the
nucleon.
The study of excitations 
is hoped to provide a clue to the intra-nucleon/baryon interactions,
in particular the degrees of freedom effectively at work.
These are not necessarily just quarks and gluons mediating the 
colour force between them.
Due to the closeness of the chiral symmetry breaking scale to the
nucleon mass/size scale, the associated Goldstone bosons also enter
as effective ``elementary'' objects \cite{GR96,MG84}.
It therefore does not come as a surprise that in some aspects, 
models which include interactions of 
the light mesons with quarks are more successful 
than genuine three quark models with pure colour interactions in
for example, the parity ordering of the lowest nucleon excitations \cite{GR96}.
Meson-baryon interactions appear to play an important role  
in baryon excitations \cite{Dalitz,SW88,KWW97,G-RLN04,LK04, Borasoy07, Bruns11, Oller01, Borasoy06}.  
Some of the states which persistingly resisted a conventionial 
three-quark explanation, for example the $(1440)1/2^+$ ``Roper resonance'' 
or the $\Lambda(1405)$, 
are likely ``dynamically'' generated through meson-baryon interactions
at least to some extent.
In addition to the interaction of pseudoscalar mesons with baryons, 
vector mesons should also contribute to dynamic resonance formation 
\cite{GOV09,Sarkar10,OR10}. 
Degenerate states of $J^P = 1/2^-,3/2^-$ are then expected, 
in particular in the mass region around 2 GeV \cite{Oset11}.

Recently, such states may have been found  
in the photoproduction reaction $\gamma + p \rightarrow K^0_S \Sigma^+$
\cite{Ewald11,Ramos-Oset13}.
A rapid fall of the cross section with increasing energy is observed 
in the vicinity of the $K^*\Lambda/\Sigma^0$ thresholds,  
changing from forward peaked to a flat angular distribution.
The effect is strong enough to generate a cusp-like structure
in the total cross section.
In Ref.\,\cite{Ewald11} this is discussed as the possible 
changeover from a $t$-channel
mechanism in $K^0$ photoproduction to the formation of an intermediate 
$s$-channel with $L = 0$ internal angular momentum formed by a 
$K^*$ vector meson interacting with an intermediate $\Lambda$ or $\Sigma$ 
hyperon. 

In order to further investigate the reaction mechanism between
$K^0$ and $K^*$ thresholds,
the analysis of the data reported in 
Ref.\,\cite{Ewald11} was extended.
In addition to the unpolarised cross sections, the photon beam asymmetry
and the hyperon recoil polarisation were also extracted \cite{Ewald10}.
This paper is organised as follows: The next section gives a brief description 
of the experiment.
Sections \ref{sec:BeamAsymmetry} and \ref{sec:RecoilPolarisation} describe the extraction of beam asymmetry and recoil polarisation. 
The results are then discussed in section \ref{sec:Discussion}, and
the paper concludes with a summary and outlook.

\section{Experiment}
\label{sec:Experiment}

Using the combined Crystal Barrel \cite{Aker92} 
and TAPS \cite{Novotny91,Gabler94} detector system,
the experiment was carried out at the Electron Stretcher Accelerator
ELSA \cite{Hillert06} of the University of Bonn's Physikalisches Institut. 
At an electron beam energy of $E_0 = 3.2$ GeV, tagged photon beams
were generated by coherent bremsstrahlung from a $500\,\mu$m thick 
diamond radiator.
Linear polarisation is obtained within the coherent intensity peaks.
The plane of linear polarisation and the energy of the coherent
peaks were both chosen through the orientation of the
radiator crystal relative to the electron beam by means of a commercial
goniometer.
The coherent peaks were set at photon energies of
$E_\gamma = 1305$, $1515$, and $1610$~MeV, with
maximum photon polarisations of $P_\gamma = 0.49$, $0.42$, and $0.39$ respectively.
The method of coherent bremsstrahlung and the performance of the setup are 
described in detail in Ref.\,\cite{Elsner09}.

The bremsstrahlung electrons were momentum analysed in
a magnetic ``tagging'' spectrometer, using a 480 channel 
scintillating fibre detector at high electron energies 
(corresponding to low photon energies, i.e. high rates),
and a MWPC at low electron energies, i.e. low rates.
A photon energy range of $E_\gamma = 0.18$--$0.92 E_0$ was covered
with an energy resolution between 10 and 25\,MeV, 
depending on the energy of the tagging electron.
Accurate tagger 
timing information was provided by 14 slightly overlapping scintillator bars.
The tagging system was run at electron rates up to $10^7$\,Hz.
The absolute photon flux was measured \,\cite{Ewald11},
but practically cancelled out in the polarisation observables
presented here.

The photon beam impinged upon a liquid hydrogen target contained 
in a $5.3$ cm long cell with 80\,$\mu$m Kapton windows. 
The reaction products were observed in the Crystal Barrel and TAPS
spectrometers, augmented by  
a cylindrical three layer scintillating fibre detector \cite{Suft05}
inside the barrel. 
In total, the detector system covered a polar angular range of
$5.8$ -- 165 degrees.
Further details of setup and readout are given, for example, in
Refs.\,\cite{Gutz14, Elsner07}.  

The detector setup is ideally suited for multi-photonic final states.
Therefore, the $K^0 \Sigma^+$ reaction was investigated in the neutral
decay modes $K^0_s \rightarrow \pi^0 \pi^0$ (B.R. $31.4\,\%$)
and $\Sigma^+ \rightarrow p \pi^0$ (B.R. $51.6\,\%$),
yielding 6 photons and the proton.
Event selection and data analysis were done as described in 
\cite{Ewald11}.
%
Here we concentrate on the aspects which are important to extract the
polarisation observables.

\section{Photon beam asymmetry}
\label{sec:BeamAsymmetry}

\begin{center}
\begin{figure}
\resizebox{0.5\textwidth}{!}{%
\includegraphics[clip=true]{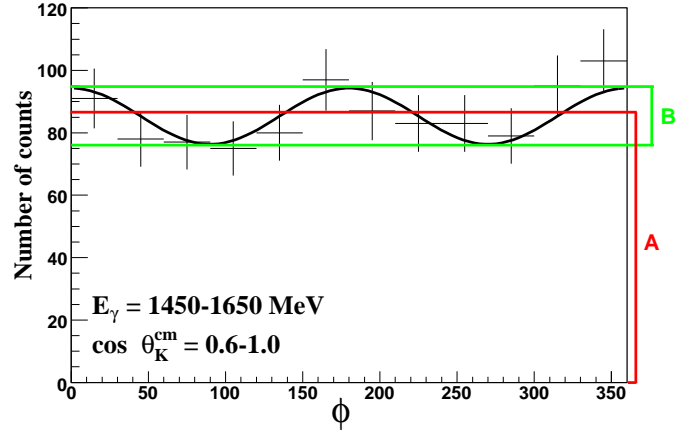}
}
\caption{Example of the azimuthal modulation of the $K^0$ yield in one bin (interval inset).
From a fit of the function $f(\phi) = A (1 - \frac{B}{A} \cos 2\phi)$ the
product $P_{lin}\Sigma_\gamma$ is determined as described in the text.}
\label{fig:azimuthalModulation}    
\end{figure}
\end{center}

Accounting for a linearly polarised photon beam, 
the cross section of photoproduction of pseudoscalar mesons off a nucleon
can be written in the form \cite{KDT95}
\begin{equation}
\frac{d\sigma}{d\Omega} = \left( \frac{d\sigma}{d\Omega} \right)_0
                  \left( 1 - P_{lin}\Sigma_\gamma \cos 2\phi \right),
\label{eq:x-sec_photoproduction}
\end{equation}
where $(d\sigma/d\Omega)_0$ is the polarisation independent cross section,
$P_{lin}$ the degree of linear polarisation, and $\Sigma_\gamma$
the photon beam asymmetry.
The product $P_{lin}\Sigma_\gamma$ determines the magnitude of modulation
of the cross section with the azimuthal angle $\phi$ beween the plane
of linear polarisation and the ejected meson.
This $\phi$-angular modulation was determined as described in 
Ref.\,\cite{Elsner07} by separate fits of the function
\begin{equation}
f(\phi) = A \left( 1 - \frac{B}{A} \cos 2\phi) \right)
\label{eq:Fit}
\end{equation}
to the $K^0$ yield in three bins of photon energy 
(1050 -- 1250, 1250 -- 1450, and 1450 -- 1650\,MeV) and five bins in the  
centre-of-mass polar angle of the $K^0$ ($\theta_{K}^{cm}$), each
$0.4$-wide in $\cos\theta_{K}^{cm}$.
An example is shown in Fig.\,\ref{fig:azimuthalModulation}.
%
The beam asymmetry, $\Sigma_\gamma$, can be extracted once the degree of 
beam polarisation is determined from the form of the 
tagged electron spectrum as also described in Ref.\,\cite{Elsner07}. 

While unpolarised background leaves the absolute amplitude of the 
azimuthal modulations unchanged, it affects the relative
strength through the denominator of $B/A$ in Eq.\,\ref{eq:Fit}.
Hence, background subtraction in the azimuthal yield spectra is
as important as for cross section measurements and was performed 
as in \cite{Ewald11}.
Polarised background would even modify the angular modulation.
Therefore, the dominating background channel of $2 \pi^0$ production was 
investigated in this respect. 
It did not show any azimuthal asymmetries for invariant masses close to the $K^0$
mass distribution, so no correction of the absolute magnitude of the observed modulation
was necessary.

The extracted azimuthal asymmetries are generally susceptible to detector
(and/or analysis) inefficiencies which vary with $\phi$.
Such instrumental effects were extensively studied in 
Refs. \cite{Elsner07, Klein08} and were investigated in the same way for the $K^0\Sigma^+$ channel: 
A pure $\cos 2\phi$ distribution 
as is expected from the reaction cross section
carries redundant information in the two intervals 
$\phi = [0,\pi]$ and $[\pi,2\pi]$.
Therefore fitted individually, both intervals are expected to yield
the same result for the fit parameters $A$ and $B$ of 
Eq.\,\ref{eq:Fit}.
Deviations are taken as contributions to the systematic error.
To avoid large statistical fluctuations in these deviations, the
systematic error contribution is averaged over each energy bin with the 
adjacent bins and weighted according to the statistical error. Uncertainties 
in background subtraction were found to be less important and
uncertainties in the beam polarisation was determined as 2\%~\cite{Elsner09},
which was considered negligible to the extracted asymmetries. 

\begin{figure}
\resizebox{0.5\textwidth}{!}{%
\includegraphics[trim=0cm 0cm 0cm 0cm, clip=true]{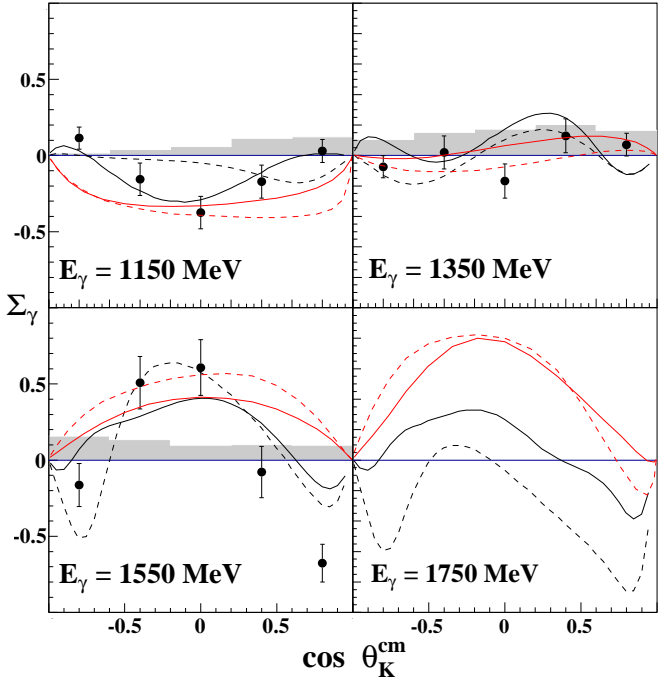}
}
\caption{Angular distribution of the photon beam asymmetry 
$\Sigma_\gamma$ in the three bins of photon energy indicated in the diagrams. 
The error bars attached to the data points are purely statistical, 
the systematic errors are indicated by the grey bars on the abscissa. 
Curves represent the results of the Bonn-Gatchina-PWA \cite{PWA_Basic}
solutions, BG2011-02m (black dashed) and BG2011-02 (black solid), 
and the K-MAID \cite{K-MAID_webseite} parametrisations,
standard (red solid) and modified as in Ref.\,\cite{Ewald11} 
to study the origin of the cross section anomaly at the $K^*$ threshold 
(red dashed).
The fourth energy bin is added to show the behaviour of the parametrisations 
across the $K^*$ thresholds where there is no data yet available.
}
\label{fig:BeamAsymmetry}    
\end{figure}

The results for the beam asymmetries are shown in Fig.\,\ref{fig:BeamAsymmetry}.
Attached to the data points are the statistical errors.
The grey bars on the abscissa indicate the systematic errors 
which are obtained by adding the individual contributions in quadrature.

The curves in Fig.\,\ref{fig:BeamAsymmetry} show parametrisations of
$\gamma\,p \rightarrow K^0_S\,\Sigma^+$ photoproduction.
The Bonn-Gatchina PWA \cite{PWA_Basic} and K-MAID \cite{K-MAID_webseite} 
are both represented in two versions: The Bonn-Gatchina 
solution BG2011-02 (black solid) includes $\gamma\,p \rightarrow K^0_S\,\Sigma^+$  
recoil polarisation data of previous CBELSA/TAPS measurements \cite{Castelijns08},
however no photon asymmetry data.
BG2011-02m (black dashed, see Ref. \cite{Gutz14} for a detailed description) is an improved variant of the Bonn-Gatchina 
solution BG2011-02, including our beam asymmetry data in Fig.\,\ref{fig:BeamAsymmetry} and new CLAS
 $\gamma\,p \rightarrow K^0_S\,\Sigma^+$ recoil polarisation data \cite{Napali14}.
K-MAID with standard parameters is shown as the solid red line, 
and with `modified' parameters in red.
The modification is discussed in Ref.\,\cite{Ewald11}.
Essentially, the $K^*$ $t$-channel exchange was switched off 
to study the effect on the cross section 
in the region of the cusp-like anomaly at the $K^*$ threshold,
cf. \cite{Ewald11}.
A (re-) fit of the data was not attempted with K-MAID,
neither in the standard nor the modified version.

\begin{figure}
\resizebox{0.5\textwidth}{!}{%
\includegraphics[clip=true]{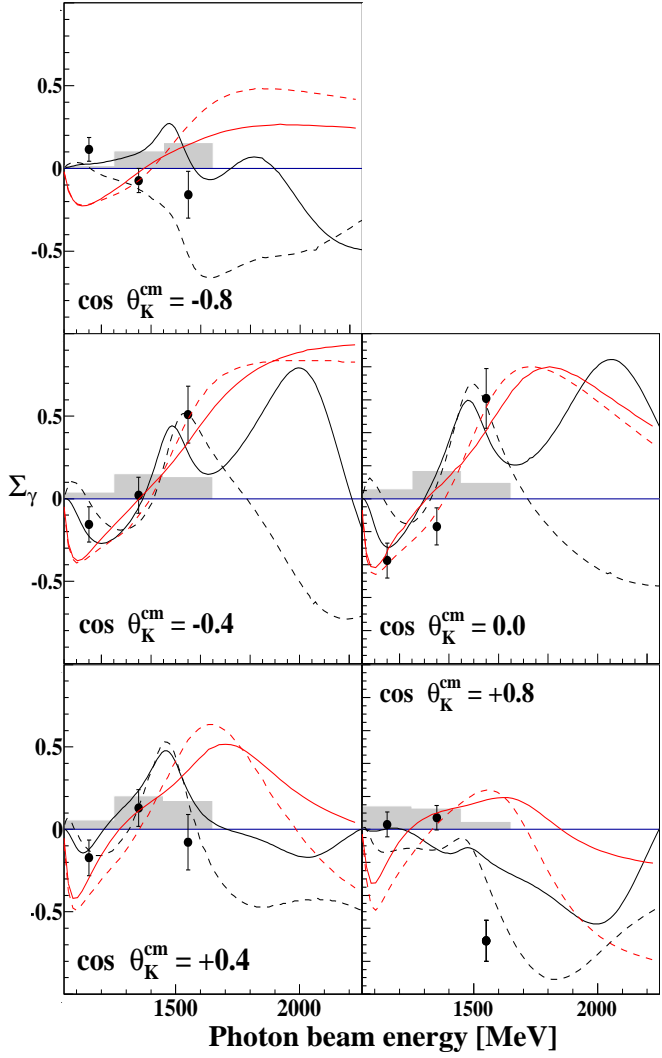}
}
\caption{Energy dependence the photon beam asymmetry 
$\Sigma_\gamma$ in the measured five bins of $\cos\,\theta_K^{cm}$.
Errors and curves are as in Fig.\,\ref{fig:BeamAsymmetry}. 
}
\label{fig:BeamAsymmetry_EnergyDependence}    
\end{figure}

Our measurement is the first one yet of the photon beam
asymmetry in this reaction channel.
Therefore, no comparison to other data can be made in 
Fig.\,\ref{fig:BeamAsymmetry}.
The data show interesting behaviour.
At threshold the photon beam asymmetry is negative\footnote{note 
that it is bounded to zero at $|\cos \theta| = 1$},
then compatible with zero throughout the intermediate energy bin.
At higher energies the beam asymmetry changes sign and turns clearly
positive, except at forward directions where it becomes strongly
negative.   
The highest energy bin shown in Fig.\,\ref{fig:BeamAsymmetry}  
illustrates the behaviour of the parameterisations 
across the $K^*$ thresholds, which are at
$E_\gamma = 1678.2$\,MeV ($E_{cm} = 2007.4$ MeV) 
for the $K^{*+}\Lambda$ final state, 
and at $E_\gamma = 1848.1$ MeV ($E_{cm} = 2085.5$\,MeV) for $K^{*0}\Sigma^+$.
Unfortunately, there are no data yet.
The same data of Fig.\,\ref{fig:BeamAsymmetry} is presented
in dependence of the photon beam energy, $E_\gamma$, in
Fig.\,\ref{fig:BeamAsymmetry_EnergyDependence}.

\section{Recoil polarisation}
\label{sec:RecoilPolarisation}

In meson photoproduction the recoiling baryons generally carry
polarisation. 
With linearly polarised photon beams,
$\sin 2\varphi$ and $\cos 2\varphi$ modulations of the recoil 
polarisation are obtained \cite{DT92}. 
The angle $\varphi$ denotes the azimuth between the plane of linear 
polarisation and the reaction plane
which is spanned by the ejected kaons and hyperons.
Due to the lack of statistics however, 
in the present analysis all relative orientations 
of polarisation and reaction planes were integrated over.
This, effectively, corresponds to unpolarised photons, 
in which case only one polarisation component remains non-zero.
This is usually called the recoil polarisation, $\vec P$, and 
due to parity conservation it is oriented normal to the
reaction plane.
   
In the studied reaction, the weak decay of the final state $\Sigma^+$ 
enables the reconstruction of the magnitude of its recoil polarisation, $P$.
The decay angular distribution has the form
\begin{equation}
W(\theta_p) = \frac{1}{2} \left( 1 + \alpha_0 P\, \cos\,\theta_p \right)
\label{eq:DAD}
\end{equation}
with $\alpha_0$ denoting the so-called decay parameter,
and $\theta_p$ the angle between the decay proton direction
and the normal of the reaction plane. 
Consequently, a count rate asymmetry 
\begin{equation}
\frac{N_\uparrow - N_\downarrow}{N_\uparrow + N_\downarrow} = 
\frac{1}{2}\,\alpha_0\,P
\label{eq:decay-asymmetry}
\end{equation}
is obtained relative to the reaction plane, where
$N_\uparrow$ and $N_\downarrow$ represent the event numbers above and
below, respectively.
A particular benefit of the $\Sigma^+ \rightarrow \pi^0 p$ decay observed
in our experiment is the large decay parameter $\alpha_0 = - 0.980$ 
\cite{PDG}.
According to Eq.\,\ref{eq:decay-asymmetry} it results in large
asymmetries from which the recoil polarisation was then determined.
\begin{figure}
\resizebox{0.5\textwidth}{!}{%
\includegraphics[clip=true]{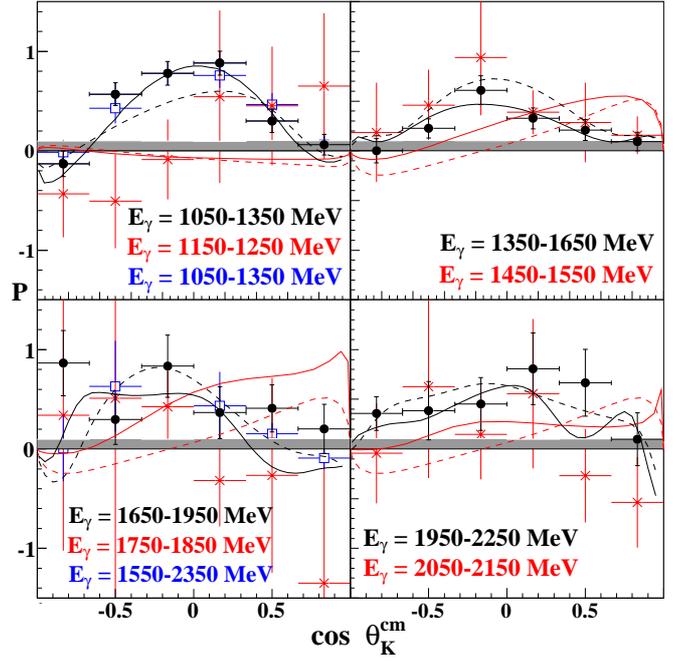}
}
\caption{\label{fig:Recoil-CB}Recoil polarisation of the $\Sigma^+$ in the 
four bins of photon energy indicated in the diagrams. 
The results of the present measurement (black dots) are compared 
to the previous CBELSA/TAPS (red crosses) \cite{Castelijns08}
and SAPHIR \cite{Lawall05} (blue squares) data. 
The latter data sets have different binning as indicated. 
Vertical bars represent the statistical errors, 
the horizontal lines attached to our data points are the
the bin widths in $\cos\theta_{K}^{cm}$.
The systematic errors are indicated by the grey bars on the abscissa. 
  Curves are the same as in Fig.\,\ref{fig:BeamAsymmetry}.}
\end{figure}

In Fig.\,\ref{fig:Recoil-CB} our results are compared to previous 
measurements of the CBELSA/TAPS collaboration, 
where a similar detector setup but unpolarised beam was used
\cite{Castelijns08}, and of SAPHIR \cite{Lawall05}.
The errors attached to the data points are purely statistical,
the shaded bands on the abscissa give an estimate of the systematic 
uncertainties.
Since, according to Eq.\,\ref{eq:decay-asymmetry}, the recoil polarisation
is determined from a ratio of event rates, some of the systematic effects
cancel out which may affect cross section or beam asymmetry measurements.
Among those are photon flux, detector inefficiencies and beam polarisation.
Remaining systematic errors were studied by variations of the cuts
applied in the analysis as described in Ref.\,\cite{Ewald11}. 

The SAPHIR data shown in Fig.\,\ref{fig:Recoil-CB} 
have different binning from our experiment, as stated in the figure. 
The SAPHIR results are compared in the bins where the weighted 
mean energy is closest.  The previous measurements of the CBELSA/TAPS collaboration 
binned the data into finer intervals.  The data shown  in Fig.\,\ref{fig:Recoil-CB}
have the same mean energies as this new data and the energy intervals are given in the
figure.
In general the data sets agree fairly well, with the older CBELSA/TAPS 
data appearing slightly low in comparison, but still within errors.
The errors of the present measurement represent a significant improvement,
in particular at lower energies.
In some energy regions this is also partly due to the use of coherent 
bremsstrahlung.
Even though the beam polarisation was integrated over, 
the coherent peaks still yielded a differential increase of photon flux.

\section{Discussion}
\label{sec:Discussion}

The photon beam asymmetry and recoil polarisation are both
observables indispensible to extract the 
reaction amplitudes \cite{CT97}, and hence the partial wave amplitudes, 
in a reliable manner.
The curves in 
Figs.~\ref{fig:BeamAsymmetry}--\ref{fig:Recoil-CB}
 demonstrate the level of agreement with the present polarisation data
which can be obtained by K-MAID \cite{K-MAID_webseite}
and the Bonn-Gatchina coupled channels PWA \cite{PWA_Basic}.

Throughout the measured kinematic range neither version of K-MAID 
reproduces the recoil polarisation data. 
There is better agreement with the PWA solution BG2011-02m in particular,  which, in contrast to K-MAID, used the 
present data of recoil polarisation and beam asymmetry as input for the fits.
Through interferences, the recoil polarisation is very sensitive 
to even small partial wave contributions. 
Hence, the observed discrepancies point to a yet incomplete resonance 
basis in K-MAID. 

The general features of the new beam asymmetry data are
reasonably described by the different parameterisations. 
Both the $K^0 \Sigma^+$ recoil polarisation and beam asymmetry data
are important inputs for  the Bonn-Gatchina PWA to
find a significant $3/2^-$ partial wave over a large energy
range. 
The fractional contributions of negative and positive 
parity states still depend on the specific PWA solution, 
but the pole structure remains nearly stable.

\begin{figure}
\resizebox{0.5\textwidth}{!}{%
\includegraphics[clip=true]{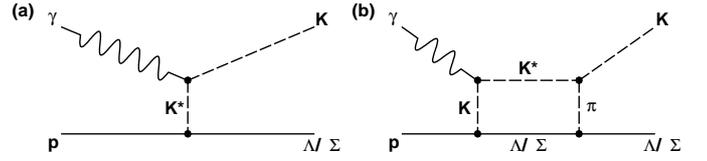}
}
\caption{$t$-exchange diagrams for $K^0$ photoproduction with an
intermediate $K^*$ (a), and an intermediate $K^0$ and pion rescattering
through subthreshold $K^*$ decay (b).  
}
\label{fig:Feynman-Diagrams}    
\end{figure}

At a mass of approximately 1880\,MeV the PWA indicates a doublet of 
negative parity $J^P = 1/2^-$ and $3/2^-$ states \cite{PWA_NegParity-PLB},
as expected by chiral symmetry restoration in high mass states
\cite{Glozman07}. 
These two nucleon resonances can be interpreted as partners
of the $J^P = 1/2^+$ and $3/2^+$ positive parity doublet 
at nearly the same mass or, alternatively, as members of the SU(6) 
56-plet expected in the third excitation band of the nucleon 
\cite{PWA_NegParity-PLB,Klempt-Metsch12}.
At higher energies in the regime of the $K^*$ 
thresholds 
the $1/2^-$ and, especially, $3/2^-$  partial waves still appear significant. 
This is also expected for the reaction mechanism hypothesised
in Ref.\,\cite{Ewald11}, where the intermediate $K^*$ and $\Lambda$ or
$\Sigma$ couple in an S-wave (Fig.\,\ref{fig:Feynman-Diagrams}b) 
to form a quasi-bound state.

While the parametrisations of the photon beam asymmetry 
agree reasonably well with our data at intermediate angles,
the situation is worse in forward directions.
The highest energy data in the forward bins in particular, appear
more negative than expected
 (Fig.\,\ref{fig:BeamAsymmetry_EnergyDependence} bottom right and
Fig.\,\ref{fig:BeamAsymmetry} bottom left).
This is interesting, because
it is the region just below the
cusp-like structure observed in the forward cross section \cite{Ewald11}.

Close to the $K^*$ threshold, quasibound $K^*$-hyperon states are
expected  in chiral unitary approaches  
through the interaction of the nonet of vector mesons with the octet
of baryons \cite{OR10}.  If a quasi-bound vector meson-hyperon state was formed, the
beam asymmetry would be expected to show structures which are sensitive 
to this.  No model calculations have yet investigated the effect on the beam asymmetry in detail.
However, there are recent calculations~\cite{Ramos-Oset13} to study the observed cross section anomaly in $K^0\Sigma^+$ 
photoproduction \cite{Ewald11}.
Subthreshold $K^*$ production and subsequent
rescattering may be responsible for the strong downturn of the 
$K^0\Sigma^+$ cross section at the $K^*$ thresholds. 
The effect appears associated with a delicate interference between 
$K^*\Lambda$ and 
$K^*\Sigma$ intermediate states which is found strongly destructive 
off the proton, but much less off the neutron.
This offers a way to test the model.
Our new data, in particular the photon asymmetry, provide a further 
testing ground for such models. 

It is desirable to further extend 
the data base.
The energy range of the presented beam asymmetry data is 
still restricted to somewhat below the $K^*$ thresholds.
To shed more light on the reaction mechanism and the possible formation
of a dynamically generated vector meson-baryon state,
it will be mandatory to extend measurements over and beyond the $K^*$ 
thresholds, where the PWA solutions have the largest discrepancies.

\section{Summary and outlook}
\label{sec:Conclusions}

Single polarisation observables in $K^0\Sigma^+$ photoproduction
off the proton were measured.
The recoil polarisation, $P$, was determined from threshold to 
$E_\gamma = 2250$\,MeV, agreeing
well with previous measurements.
While the description by the K-MAID parametrisation 
without attempting to fit the new data
remains unsatisfactory, 
new fits bring the Bonn-Gatchina 
PWA into good agreement with the measured recoil polarisations 
\cite{PWA_NegParity-PLB}.

The photon beam asymmetry was measured for the first time in
$K^0\Sigma^+$ photoproduction.
The Bonn-Gatchina PWA, as well as the original and a modified version of 
K-MAID describe the data fairly well in the intermediate angular range.
In the most forward direction the beam asymmetry shows the interesting feature 
that it turns strongly negative just below the $K^*$ threshold,
where a strong decline of the forward cross section was observed 
\cite{Ewald11}.
A recent calculation in the chiral unitary framework indicates
that this effect may be related to a dynamically generated 
vector meson-hyperon state \cite{Ramos-Oset13}.
Whether this model is able to reproduce the polarisation
data remains to be seen.
 
Definite conclusions on the reaction mechanism in the cusp region
will require the beam asymmetry to be 
measured across the $K^*$ threshold. 
In addition, beam-target and beam-recoil double polarisation observables
will be necessary to reveal the helicity structure in that energy regime. 
Such measurements will be subject to future investigations
using the CBELSA/TAPS and particularly the new BGO-OD detector setup 
\cite{bantes13, HS10-BGOOD, Schmieden09} at ELSA.

\section*{Acknowledgements}

We thank the staff and shift-students of the ELSA accelerator for
providing an excellent beam.
This work was supported by the federal state of 
{\em North-Rhine Westphalia}, the
{\em Deutsche Forschungsgemeinschaft} within the SFB/TR-16 and the
{\em Schweizerischer Nationalfonds}.

%
%


\begin{thebibliography}{}
%
%

\bibitem{GR96}
L.Ya. Glozman and D.O. Riska,
Physics Reports {\bf 268} (1996) 263

\bibitem{MG84}
A. Manohar and H. Georgi,
Nucl. Phys. {\bf B 234} (1984) 189

\bibitem{Dalitz}
R.H. Dalitz and J.G. McGinley, 
in {\em Low and Intermediate Energy Kaon-Nucleon Physics}, ed. by 
E. Ferrari and G. Violini, Reidel, Boston (1981) 381;
R. H. Dalitz, T.C. Wong, and G. Rajasekaran,
Phys. Rev {\bf 153} (1967) 1617

\bibitem{SW88}
P.B. Siegel, and W. Weise,
Phys. Rev {\bf C 38} (1988) 2221

\bibitem{KWW97}
N. Kaiser, T. Waas, and W. Weise,
Nucl. Phys, {\bf A 612} (1997) 297

\bibitem{G-RLN04}
C. Garcia-Recio, M.F.M. Lutz, and J. Nieves,
Phys. Lett {\bf B 582} (2004) 49 

\bibitem{LK04}
M.F.M. Lutz and E.E. Kolomeitsev,
Phys. Lett. {\bf B 585} (2004) 243


\bibitem{Borasoy07}
B.~Borasoy, P.C.~Bruns, U.-G.~Mei\ss ner and R.~Ni\ss ler,
Eur. Phys. J. {\bf A 34} (2007) 161

\bibitem{Bruns11}
P.C.~Bruns,  M.~Mai and U.-G.~Mei\ss ner,
Phys. Lett. {\bf B 697} (2011) 254

\bibitem{Oller01}
J.A.~Oller and U.-G.~Mei\ss ner,
Phys. Lett. {\bf B 500} (2001) 263

\bibitem{Borasoy06}
B.~Borasoy, U.-G.~Mei\ss ner and R.~Ni\ss ler,  
Phys. Rev. {\bf C 74} (2006) 055201



\bibitem{GOV09}
P. Gonzalez, E. Oset and J. Vijande,
Phys. Rev {\bf C 79} (2009) 025209

\bibitem{Sarkar10}
S. Sarkar et al.,
Eur. Phys. J. {\bf A 44} (2010) 431

\bibitem{OR10}
E. Oset and A. Ramos,
Eur. Phys. J. {\bf A 44} (2010) 445

\bibitem{Oset11}
E. Oset et al., AIP Conf.Proc. 1388 (2011) 295,
arXiv:1103.0807v1 [nucl-th]

\bibitem{Ewald11}
R. Ewald et al.,
Phys. Lett. {\bf B 713} (2012) 180

\bibitem{Ramos-Oset13}
  A.~Ramos and E.~Oset,
  Phys.\ Lett. {\bf B 727}, (2013) 287

\bibitem{Ewald10}
R. Ewald, Doctoral Thesis, Bonn (2010) 

\bibitem{Aker92}
E. Aker et al.,
Nucl. Instrum. Methods {\bf A 321} (1992) 69

\bibitem{Novotny91}
R. Novotny et al.,
IEEE Trans. Nucl. Sci. {\bf 38} (1991) 379 

\bibitem{Gabler94}
A.R. Gabler et al.,
Nucl. Instrum. Methods {\bf A 346} (1994) 168

\bibitem{Hillert06}
W. Hillert,
Eur. Phys. J. {\bf A 28}, s01 (2006) 139 

\bibitem{Elsner09}
D. Elsner et al.,
Eur. Phys. J. {\bf A 39} (2009) 373 

\bibitem{Suft05}
G. Suft et al.,
Nucl. Instrum. Methods {\bf A 538} (2005) 416

\bibitem{Gutz14}
E. Gutz, V. Crede, V. Sokhoyan, H. van Pee et al.,
Submitted to Eur.\ Phys.\ J.\  {\bf A},
arXiv:1402.4125v1 [nucl-ex] (2014)

\bibitem{Elsner07}
D. Elsner et al.,
Eur.\ Phys.\ J.\  {\bf A33} (2007) 147

\bibitem{KDT95}
G. Kn\"ochlein, D. Drechsel and L. Tiator, 
Z. Phys. \textbf{A352} (1995) 327

\bibitem{Klein08}
Frank Klein et al., 
Phys. Rev {\bf D 78} (2008) 117101

\bibitem{PWA_Basic}
A.V. Anisovich et al.,
Eur. Phys. J. {\bf A 47} (2011) 27, and references therein

\bibitem{K-MAID_webseite}
D. Drechsel et al., http://www.kph.uni-mainz.de/MAID/  (Version 29.3.2007)


\bibitem{Castelijns08}
R. Castelijns et al.,
Eur. Phys. J. {\bf A 35} (2008) 39

\bibitem{Napali14}
C.S. Napali et al. (CLAS Collaboration), Phys. Rev. {\bf C 87} (2013) 045206

\bibitem{DT92} 
D. Drechsel and L. Tiator,
Journal of Physics {\bf G 18} (1992) 449

\bibitem{PDG} 
K. Nakamura et al. (Particle Data Group),
Journal of Physics {\bf G 37} (2010) 075021

\bibitem{Lawall05} 
R. Lawall et al.,
Eur. Phys. J. {\bf A 24} (2005) 275

\bibitem{CT97} 
W.-T. Chiang and F. Tabakin, 
Phys. Rev. \textbf{C55} (1997) 2054


\bibitem{PWA_NegParity-PLB} 
A.V. Anisovich et al., 
Phys. Lett. {\bf B 711} (2012) 162

\bibitem{Glozman07} 
L.Ya. Glozman,
Physics Reports {\bf 444} (2007) 1

\bibitem{Klempt-Metsch12} 
E. Klempt and B.C. Metsch,
Eur. Phys. J. {\bf A 48} (2012) 127


\bibitem{bantes13}
B. Bantes et al., Int. J. Mod. Phys: Conf. Ser. {\bf 26} (2014) 1460093

\bibitem{HS10-BGOOD} 
H. Schmieden, Int. J. Mod. Phys. {\bf E 19} (2010) 1043

\bibitem{Schmieden09}
H. Schmieden, Chinese Physics {\bf C 33} (2009) 1146


\end{thebibliography}

\end{document}